\documentclass[pra,aps,amssymb,twocolumn,longbibliography]{revtex4-1}
\usepackage{amsmath}
\usepackage{amssymb}
\usepackage{amsthm}
\usepackage{amsfonts}
\usepackage{mathtools}
\usepackage{siunitx}
\usepackage{listings}
\usepackage{enumerate}
\usepackage{latexsym}
\usepackage{color}
\usepackage[colorlinks=true,urlcolor=blue,citecolor=blue,linkcolor=blue]{hyperref}
\usepackage{psfrag}
\usepackage{bm}

\usepackage{graphicx}
\usepackage{subfigure}
\usepackage{cleveref}

\usepackage{booktabs}

\renewcommand{\arraystretch}{1.3}
\AtBeginDocument{
	\heavyrulewidth=.08em
	\lightrulewidth=.05em
	\cmidrulewidth=.03em
	\belowrulesep=.65ex
	\belowbottomsep=0pt
	\aboverulesep=.4ex
	\abovetopsep=0pt
	\cmidrulesep=\doublerulesep
	\cmidrulekern=.5em
	\defaultaddspace=.5em
}

\newcommand{\unit}[1]{\ensuremath{~\mathrm{#1}}}
\newcommand{\sub}[1]{$_\text{#1}$}
\renewcommand{\vec}[1]{\ensuremath{\mathbf{#1}}}

\newcommand{\kdotp}{\ifmmode\vec{k}\cdot\vec{p}\else $\kdotp$ \fi}

\begin{document} 

\tolerance 10000

\title{Hidden Weyl Points in Centrosymmetric Paramagnetic Metals}

\author{Dominik Gresch$^1$}
\author{QuanSheng Wu$^1$}
\author{Georg W. Winkler$^1$}
\author{Alexey A. Soluyanov$^{1,2}$}

\affiliation{$^{1}$Theoretical Physics and Station Q Zurich, ETH Zurich, 8093 Zurich, Switzerland}
\affiliation{$^{2}$Department of Physics, St. Petersburg State University, St. Petersburg, 199034 Russia}

\date{\today}
\begin{abstract}
	The transition metal dipnictides TaAs\sub{2}, TaSb\sub{2}, NbAs\sub{2} and NbSb\sub{2} have recently sparked interest for exhibiting giant magnetoresistance. While the exact nature of magnetoresistance in these materials is still under active investigation, there are experimental results indicating anisotropic negative magnetoresistance. We study the effect of magnetic field on the band structure topology of these materials by applying a Zeeman splitting. In the absence of magnetic field, we find that the materials are weak topological insulators, which is in agreement with previous studies. When the magnetic field is applied, we find that type - II Weyl points form. This result is found first from a symmetry argument, and then numerically for a \kdotp model of TaAs\sub{2} and a tight-binding model of NbSb\sub{2}. This effect can be of help in search for an explanation of the anomalous magnetoresistance in these materials.
\end{abstract}

\maketitle

\section{Introduction}
Weyl nodes are point-like crossings of two energy bands, with a linear dispersion. Locally, they can be described by Hamiltonian of the form 
\begin{equation}\label{eqn:weyl_hamilton}
H(k) = \sum_{i, j} k_i A_{i, j} \sigma_j
\end{equation} 
where $i \in \{x, y, z\}$, $j \in \{0, x, y, z\}$. Topologically, a Weyl node can be characterized by being a quantized source or sink of Berry curvature, depending on its chirality~\cite{Volovik-JETPL87}. Due to this quantized nature, Weyl points can only be created or annihilated in pairs of opposite chirality. 

It was recently shown~\cite{Soluyanov-Nat15} that Weyl nodes come in two types. Type - I Weyl fermions have a point-like Fermi surface. When a magnetic field is applied, they exhibit a chiral Landau level~\cite{Adler-PR69, Bell-Jackiw-NC69, Nielsen-PLB83} regardless of the magnetic field direction. When this chiral Landau level crosses the Fermi level, it can be a source of reduced magneto-resistance~\cite{Abrikosov-PRB98, Son-PRB13, Huang-PRX15, Xiong-Sci15, Arnold-NatComm16, Yang-ARX15}. Type - II Weyl fermions on the other hand have an energy spectrum that is tilted by having a strong $\sigma_0$ contribution to the Hamiltonian (\cref{eqn:weyl_hamilton}). As a consequence, the Fermi surface becomes open and the chiral anomaly is anisotropic, appearing only for certain magnetic field directions.

Inversion symmetry $P$ maps a Weyl node at point $\vec{k}$ onto a Weyl node of opposite chirality at $-\vec{k}$. Similarly, time-reversal symmetry $\mathcal{T}$ maps a Weyl point at $\vec{k}$ onto one at $-\vec{k}$, but without changing its chirality. Consequently, in the presence of the product symmetry $\mathcal{T} * P$, Weyl nodes are mapped into themselves but with opposite chirality. This four-fold degenerate crossing, consisting of two superimposed Weyl points of opposite chirality, is known as a Dirac node. Unlike Weyl points, they are not protected from gapping by any quantized topological charge. Consequently, additional symmetries are needed to stabilize Dirac nodes.

In centrosymmetric non-magnetic materials, the presence of both inversion and time-reversal symmetry allows only for Dirac nodes to form. Weyl nodes are not possible unless the product symmetry $\mathcal{T} * P$ is broken. 

Recently, transition metal dipnictides of the type AB\sub{2} (A $\in$ \{Ta, Nb\}, B $\in$ \{As, Sb\}) have gained a lot of attention ~\cite{Wang-SciRep14, Wang-PRB16, Wu-APL16, Luo-ScR16, Yuan-PRB16, Li-ARX16, Shen-PRB16, Wang-ARX16} for their giant magnetoresistance. These materials are semimetals, but without a direct closure of the band gap. Consequently, they do not host any Weyl or Dirac points. 

The exact nature of magnetoresistance in these materials - especially the dependency on the direction of the magnetic field - is still under active investigation. Negative magnetoresistance has been observed experimentally for NbAs\sub{2}~\cite{Li-ARX16, Shen-PRB16}, TaAs\sub{2} ~\cite{Li-ARX16} and TaSb\sub{2} ~\cite{Li-ARX16,Li-PRB16}. Anomalous - albeit not negative - magnetoresistance has been observed for NbSb\sub{2}\cite{Wang-SciRep14} and TaAs\sub{2}~\cite{Luo-ScR16}. However, there are also experiments which point to the contrary, which is that there is no negative magnetoresistance in these materials. In Ref. \cite{Yuan-PRB16}, negative magnetoresistance was observed at first but then determined to be an artifact of the measurement setup.

In the following, we propose a mechanism for Weyl nodes to appear in these materials under the influence of a magnetic field. The chiral anomaly associated with these Weyl nodes is a possible source of negative magnetoresistance. Such an appearance of Weyl points under magnetic field has recently been proposed in Ref.~\cite{Cano-ARX16}. The mechanism with which the Weyl points appear, however, is a different one -- in this work the Weyl points appear from a previously gapped state, while the Ref.~\cite{Cano-ARX16} discusses Weyl points arising from the splitting of a four-fold crossing.

The paper is structured as follows: In the first section, the atomic and electronic structure of the four compounds is described. A four-band Hamiltonian for TaAs\sub{2} is derived from symmetry considerations and fitted to the band structure. In the second section, the topology of the band structure is studied, first without magnetic field and then by applying a Zeeman term. We find that this leads to the appearance of Weyl points.

\section{Atomic and Electronic Structure of AB$_2$ compounds}
\subsection{Atomic structure}

In the following, the atomic structure of TaAs\sub{2}~\cite{Ling-CRAScP81}, TaSb\sub{2}~\cite{Hulliger-Nat64}, NbSb\sub{2}~\cite{Lomnytska-IM05} and NbAs\sub{2}~\cite{Bensch-AC95} is described. 

The reduced unit cell of AB\sub{2} compounds has the general form
\begin{alignat}{4}
	\label{eqn:basis}
	a_1 & = (&&a,~&&b,~&&0)\\\nonumber
	a_2 & = (-&&a,~&&b,~&&0)\\\nonumber
	a_3 & = (-&&c,~&&0,~&&d)
\end{alignat}
with parameters as given in \cref{tab:uc_dimensions}~\cite{Ling-CRAScP81, Lomnytska-IM05}.

\begin{table}[h]
\begin{tabular}{@{}lllll@{}}
\toprule
 & $\mathbf{a}$ & $\mathbf{b}$ & $\mathbf{c}$ & $\mathbf{d}$\\ \midrule
\textbf{TaAs\sub{2}}~~ & $4.6655$ & $1.6915$ & $3.8420$ & $6.7330$\\
\textbf{TaSb\sub{2}} & $5.11$ & $1.822$ & $4.1950$ & $7.1502$\\
\textbf{NbAs\sub{2}} & $4.684$ & $1.698$ & $3.8309$ & $6.7933$ \\
\textbf{NbSb\sub{2}} & $5.1198$ & $1.8159$ & $4.1705$ & $7.2134$ \\ \bottomrule
\end{tabular}
\caption[]{Unit cell dimensions (in \AA) for AB\sub{2} compounds.}
\label{tab:uc_dimensions}
\end{table}

Each unit cells contains 2 formula units. The atoms are located at general Wyckoff positions $(x, -x, y), (-x, x, -y)$, for $(x, y)$ as shown in \cref{tab:atom_positions}~\cite{Ling-CRAScP81, Lomnytska-IM05}. 

\begin{figure}
	\includegraphics[width=0.8\columnwidth]{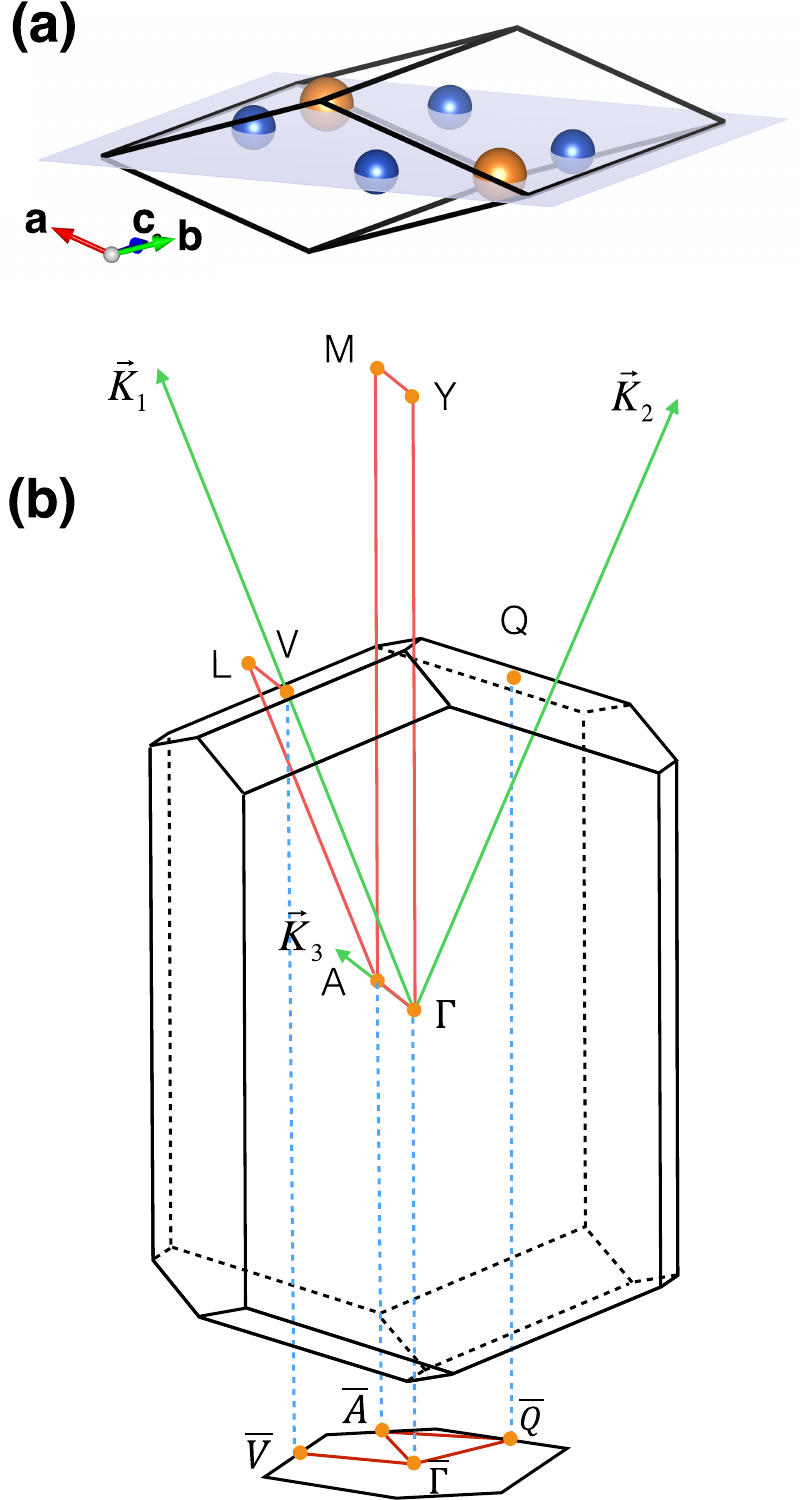}
	\caption[]{(a) Reduced unit cell of TaAs\sub{2}. (b) First BZ of TaAs\sub{2}. The k-point path and its projection onto the 010 surface are indicated.}
	\label{fig:uc_bz_TaAs2}
\end{figure}

\Cref{fig:uc_bz_TaAs2} shows the reduced unit cell and 1. BZ of TaAs\sub{2}. The k-point path along which bandstructure calculations are performed is indicated. In the basis reciprocal to that of \cref{eqn:basis}, the special k-points are given by

\begin{alignat}{3}
	&\Gamma &&= (0, 0, 0)	
	\\\nonumber
	&A &&= (0, 0, 0.5)		
	\\\nonumber
	&L &&= (0.5, 0, 0.5)		
	\\\nonumber
	&M &&= (0.5, 0.5, 0.5)		
	\\\nonumber
	&V &&= (0.5, 0, 0)		
	\\\nonumber
	&Y &&= (0.5, 0.5, 0).	
\end{alignat}

\begin{table}[h]
\begin{tabular}{@{}llll@{}}
	\toprule
	& \textbf{A} & \textbf{B1} & \textbf{B2} \\ \midrule
	\textbf{TaAs\sub{2}}~~ & $(0.157, 0.1959)$ & $(0.4054, 0.1082)$ & $(0.1389, 0.5265)$\\
	\textbf{TaSb\sub{2}} & $(0.152, 0.19)$ & $(0.405, 0.113)$ & $(0.147, 0.535)$\\
	\textbf{NbAs\sub{2}} & $(0.1574, 0.1965)$ & $(0.4059, 0.1084)$ & $(0.14, 0.528)$\\
	\textbf{NbSb\sub{2}} & $(0.1521, 0.1903)$ & $(0.4051, 0.1127)$ & $(0.1475, 0.5346)$\\ \bottomrule
\end{tabular}
\caption[]{Atomic positions $(x, y)$.}
\label{tab:atom_positions}
\end{table}

\subsection{Electronic structure}

Electronic structure calculations were performed in VASP~\cite{VASP}, with projector augmented-wave (PAW)~\cite{PAW1, PAW2} pseudopotentials. The PBE approximation ~\cite{PBE} was used, and spin-orbit coupling was included in the potentials. The self-consistent field (SCF) calculations were performed on a $11\times11\times5$ $\Gamma$ - centered grid  for TaAs\sub{2}, and a $10\times10\times5$ $\Gamma$ - centered grid for NbSb\sub{2}. The energy cutoff given in the potential files was used, which is $293.2 \unit{eV}$ for NbAs\sub{2} and NbSb\sub{2}, and $223.7\unit{eV}$ for TaAs\sub{2} and TaSb\sub{2}.

Additionally, the PBE calculations were tested against the accurate HSE06 hybrid functional~\cite{Heyd-JChPh03, HSE06}. The hybrid SCF calculations for the band structures were performed on a $\Gamma$ - centered $6\times6\times4$ grid for all materials. For the generation of the Wannier tight-binding model of NbSb\sub{2} a $\Gamma$ - centered $10\times10\times5$ grid was used.

\begin{figure}
	\includegraphics[width=\columnwidth]{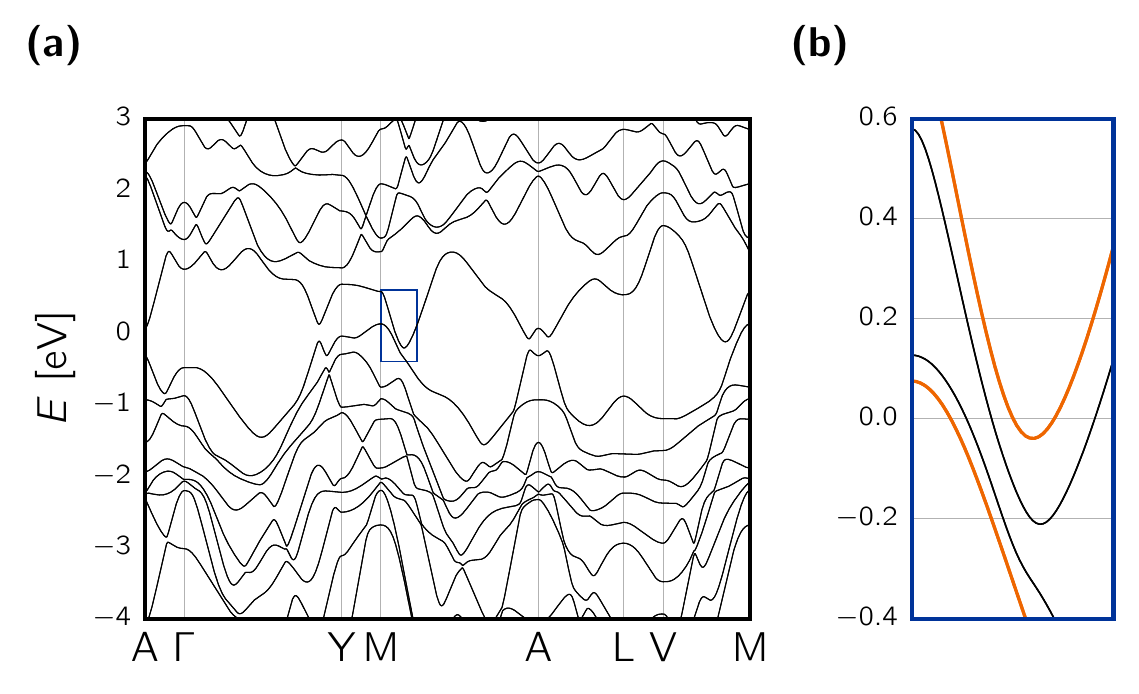}
	\includegraphics[width=\columnwidth]{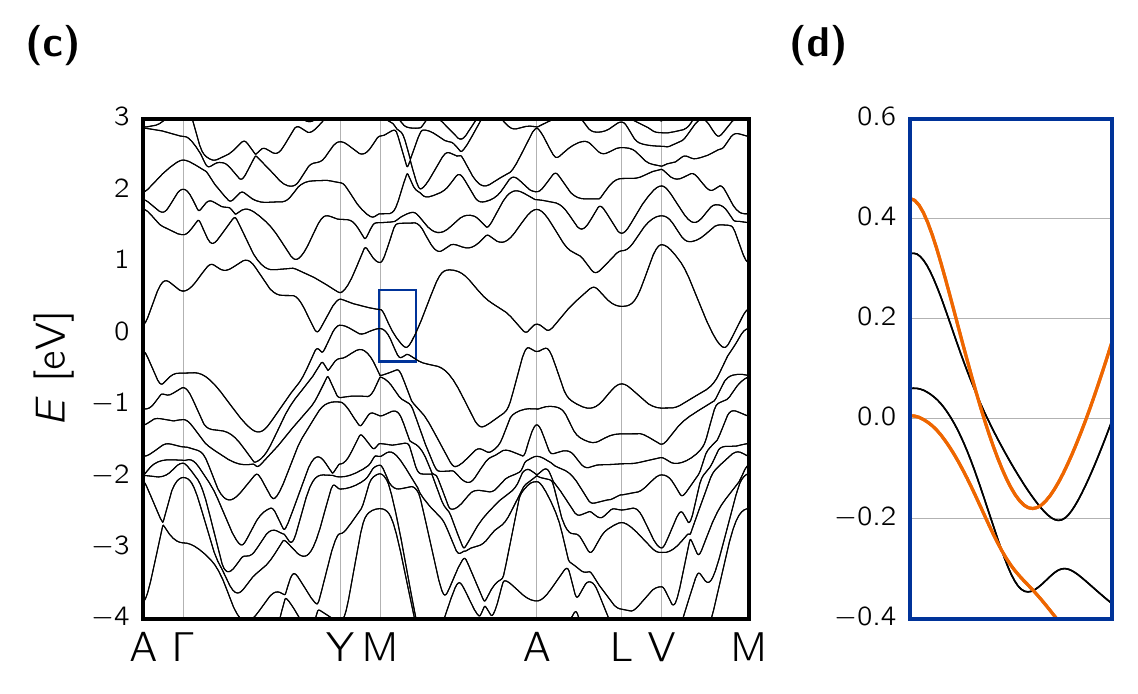}
	\includegraphics[width=\columnwidth]{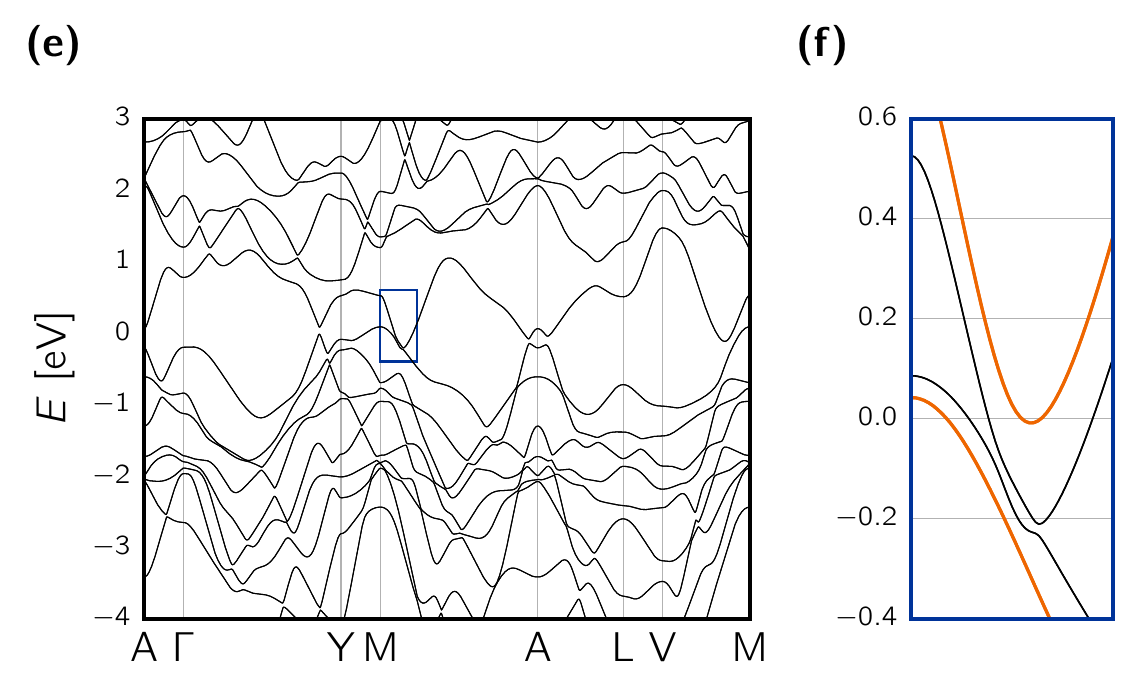}
	\includegraphics[width=\columnwidth]{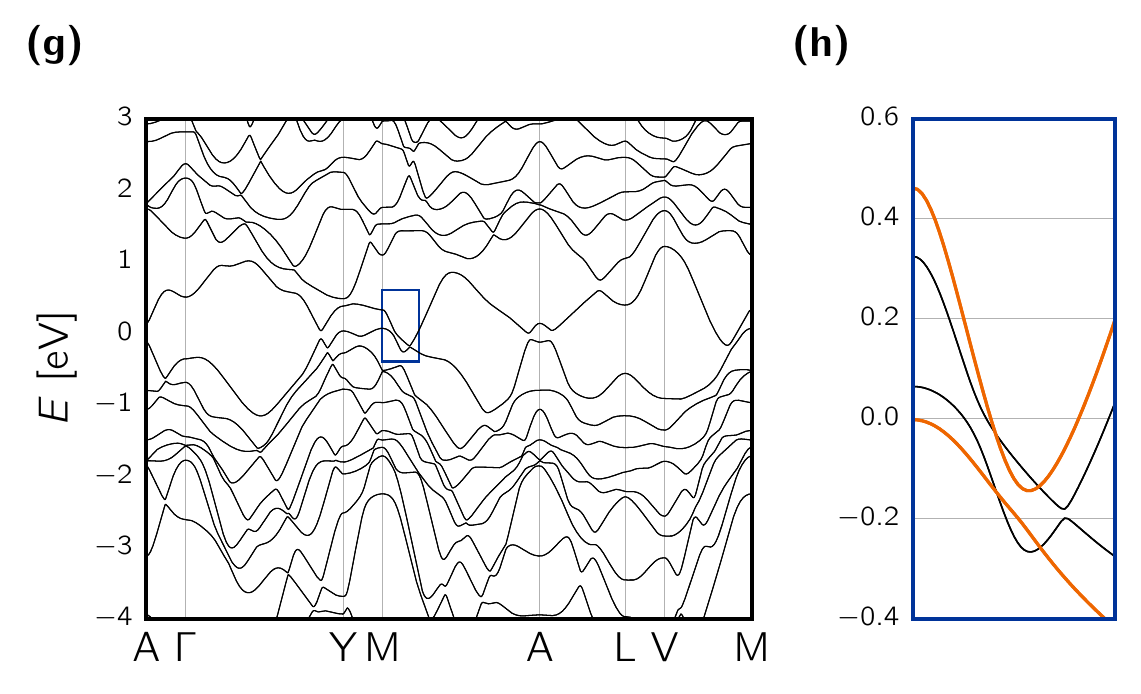}
	\caption[]{Band structures of AB\sub{2} compounds. The inset shows electron and hole pockets around $M$. The orange line represents calculations using hybrid functionals. (a-b) TaAs\sub{2} (c-d) TaSb\sub{2} (e-f) NbAs\sub{2} (g-h) NbSb\sub{2}}
	\label{fig:bands}
\end{figure}

The band structure of TaAs\sub{2} and NbSb\sub{2} is shown in \cref{fig:bands}. Both materials exhibit a pair of electron and hole pockets near the $M$ - point, where the minimum band gap is about $318 \unit{meV}$ ($120\unit{meV}$ without hybrid functionals) in the case of TaAs\sub{2}, $151 \unit{meV}$ ($98\unit{meV}$) for TaSb\sub{2}, $261 \unit{meV}$ ($22\unit{meV}$) for NbAs\sub{2}, and $67\unit{meV}$ ($18\unit{meV}$) in the case of NbSb\sub{2}. A more complete calculation of the band structure can be found for example in Ref.~\cite{Xu-PRB16}.

\subsection{Symmetry operations and $\vec{k} \cdot \vec{p}$ model}\label{ssec:kdotp}
The AB\sub{2} compounds studied here have $C2/m$ symmetry (space group 12). The rotation axis is along the cartesian $y$-axis. In reduced coordinates, the symmetry matrices are as follows:
\begingroup
\renewcommand*{\arraystretch}{1.}
\begin{alignat*}{2}
	\bullet~~ & \text{Identity} & E =  & ~\mathbb{I}_{3\times3} \\
	\bullet~~ & \text{Rotation} & C_{2y} = & 
	~\begin{pmatrix*}[r]
		0 & 1 & 0 \\
		1 & 0 & 0 \\
		0 & 0 & -1
	\end{pmatrix*} \\
	\bullet~~ & \text{Parity} & P =  & ~- \mathbb{I}_{3\times 3}\\
	\bullet~~ & \text{Mirror} & M_y = P C_{2y} =  &
	~\begin{pmatrix*}[r]
		0 & -1 & 0 \\
		-1 & 0 & 0 \\
		0 & 0 & 1
	\end{pmatrix*} 
\end{alignat*}
\endgroup

\begin{table}[h]
\begin{tabular}{lrrrr}
\toprule
& $E$ & ~$C_{2y}$ & $P$ & ~$M_y$ \\\midrule
$\mathbf{\Gamma_3^+} $ & $1$ & $~i$ & $~1$ & $i$ \\
$\mathbf{\Gamma_4^+} $ & $1$ & $-i$ & $~1$ & $-i$ \\
$\mathbf{\Gamma_3^-} $ & $1$ & $~i$ & $-1$ & ~~$-i$ \\
$\mathbf{\Gamma_4^-} $ & ~~~$1$ & ~~$-i$ & ~~$-1$ & $i$ \\
\bottomrule
\end{tabular}
\caption[]{Character table for the relevant double group representations of $C_{2m}$~\cite{Koster-book}.}
\label{tab:representations}
\end{table}

From the first-principles wave-functions, the representations corresponding to the two highest valence and two lowest conduction bands at the $M$ - point were determined using the \texttt{WIEN2k} code~\cite{w2kcode, wien2k2}. They were found to be $\Gamma_3^+, \Gamma_4^+$ and $\Gamma_3^-, \Gamma_4^-$, respectively. Their characters are shown in \cref{tab:representations}, which comes from table 15 on page 35 in Koster et al.~\cite{Koster-book}. Consequently, the symmetry representations in these four bands are given by
\begingroup
\renewcommand*{\arraystretch}{1.}
\begin{alignat*}{2}
	\bullet~~ & \text{Identity} & E = & ~\mathbb{I}_{4\times4} \\
	\bullet~~ & \text{Rotation} & C_{2y} =  & 
	~\begin{pmatrix*}[r]
	i & 0 & 0 & 0 \\
	0 & -i & 0 & 0\\
	0 & 0 & i & 0\\
	0 & 0 & 0 & -i
	\end{pmatrix*}\\
	\bullet~~ & \text{Parity} & P =  & 
	~\begin{pmatrix*}[r]
	1 & 0 & 0 & 0 \\
	0 & 1 & 0 & 0\\
	0 & 0 & -1& 0\\
	0 & 0 & 0 & -1
	\end{pmatrix*}\\
	\bullet~~ & \text{Mirror} & M_y = P C_{2y} =  & 
	~\begin{pmatrix*}[r]
	i & 0 & 0 & 0 \\
	0 & -i & 0 & 0\\
	0 & 0 &-i & 0\\
	0 & 0 & 0 & i
	\end{pmatrix*}\\
	\bullet~~ & \text{Time-reversal} & \mathcal{T} =  & 
	~\begin{pmatrix*}[r]
	0 & -1 & 0 & 0 \\
	1 & 0 & 0 & 0 \\
	0 & 0 & 0 & -1 \\
	0 & 0 & 1 & 0
	\end{pmatrix*}\hat{K}
\end{alignat*}
\endgroup

%
%
%
For each of the symmetry operations $g$, the constraint  
\begin{equation}
\mathcal{H}(\vec{k}) = D(g) \mathcal{H}(g^{-1}\vec{k})D(g^{-1})
\end{equation}
is imposed on the $4\times 4$ Hamiltonian, where $D(g)$ is the symmetry representation. By applying these constraints on the general form of a four-band Hamiltonian 
\begin{equation}
\mathcal{H}(\vec{k}) = \sum_{i,j \in \{0, x, y, z\}} C_{ij}(\vec{k}) (\sigma_i \otimes \sigma_j),
\end{equation}
we find the Hamiltonian to be of the form
\begin{align}\label{eqn:kdotp_hamilton}
\mathcal{H}(\vec{k}) =~&C_{00}(\vec{k}) (\sigma_{0} \otimes \sigma_{0}) + C_{xx}(\vec{k}) (\sigma_{x} \otimes \sigma_{x}) ~+ \\\nonumber& C_{xy}(\vec{k}) (\sigma_{x} \otimes \sigma_{y}) + C_{xz}(\vec{k}) (\sigma_{x} \otimes \sigma_{z}) ~+ \\\nonumber& C_{y0}(\vec{k}) (\sigma_{y} \otimes \sigma_{0}) + C_{z0}(\vec{k}) (\sigma_{z} \otimes \sigma_{0}),
\end{align}
where the $C_{ij}(\vec{k})$ are given up to second order in $\vec{k^*} = \vec{k} - M$ (in reduced coordinates) by
\begin{align}
C_{00}(\vec{k^*}) = ~&C_{00}^1 + C_{00}^{x^2+y^2}((k_x^*)^2 + (k_y^*)^2)~+ \\\nonumber & C_{00}^{xy} ~ k_x^*k_y^*+ C_{00}^{xz-yz} (k_x^*k_z^*- k_y^*k_z^*)~+ \\\nonumber &  C_{00}^{z^2} ~ (k_z^*)^2
\\
C_{z0}(\vec{k^*}) =~& C_{z0}^{1} + C_{z0}^{x^2+y^2} ((k_x^*)^2 + (k_y^*)^2)~+ \\\nonumber &  C_{z0}^{xy} ~ k_x^*k_y^* + C_{z0}^{xz - yz} (k_x^*k_z^* - k_y^*k_z^*)~+ \\\nonumber &  C_{z0}^{z^2} ~ (k_z^*)^2
\\
C_{xx}(\vec{k^*}) =~& C_{xx}^{x-y}(k_x^*- k_y^*) + C_{xx}^z ~ k_z^*
\\
C_{xy}(\vec{k^*}) =~& C_{xy}^{x-y}(k_x^*- k_y^*) + C_{xy}^z ~ k_z^*
\\
C_{xz}(\vec{k^*}) =~& C_{xz}^{x+y} (k_x^*+ k_y^*) 
\\
C_{y0}(\vec{k^*}) =~& C_{y0}^{x+y} (k_x^*+ k_y^*).
\end{align}

\begin{figure}
\includegraphics[width=\columnwidth]{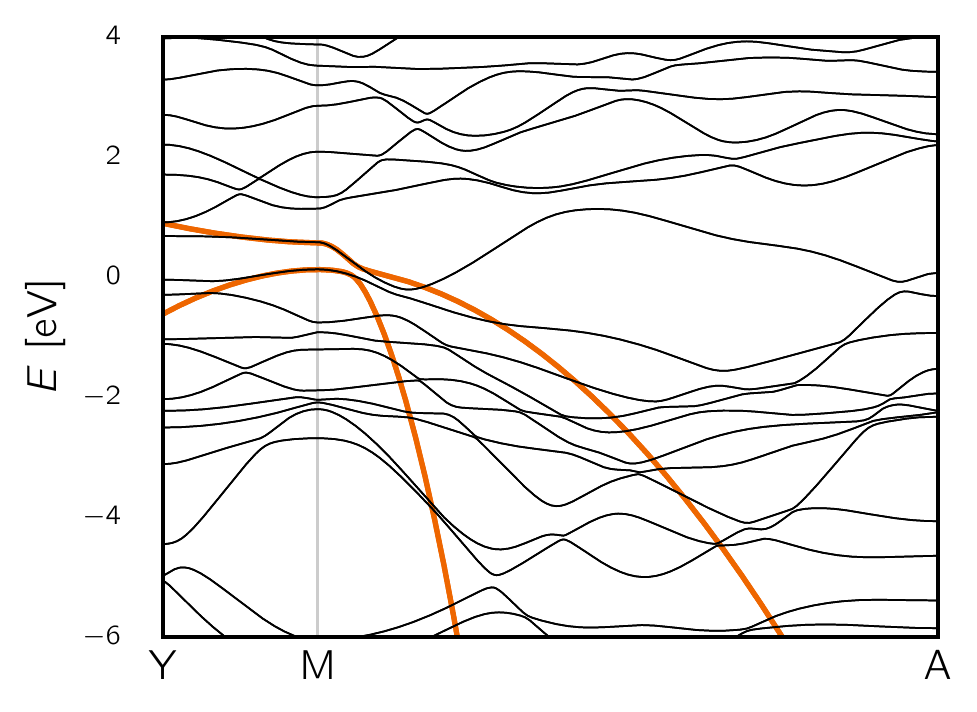}
\caption[]{TaAs\sub{2} band structure of the $\vec{k}\cdot \vec{p}$ model (thick orange line), compared to the first-principles result (black lines).}
\label{fig:bands_kdotp}
\end{figure}

These 16 parameters were numerically fitted to the band structure of TaAs\sub{2} using the \texttt{scipy}~\cite{scipy} package, to obtain the values in \cref{tab:fit_parameters}. The resulting band structure around the $M$-point is shown in \cref{fig:bands_kdotp}. Comparing it to the band structure obtained from first-principles reveals that the approximation is accurate in the immediate vicinity of the $M$-point, but breaks down at around $6 \%$ of the distance along the line $M-A$. Importantly, the minimum band gap is not preserved in this model. Nevertheless, the model can be used to qualitatively study effects in TaAs\sub{2}, owing to the fact that it contains the correct symmetry representations.

\begin{table}[h]
\begin{tabular}{@{}lllll@{}}
	\toprule
	$\bm{[}\si{\textbf{\electronvolt}} \bm{]}$ & $C^1_{00} $&$= 7.066$ & $C^1_{z0} $&$= -0.224$ \\ \midrule
	$\bm{[}\si{\textbf{\electronvolt \angstrom}} \bm{]}$ & $C^{x+y}_{xz} $&$= 1.272$ & $C^{x+y}_{y0} $&$= 1.270$ \\
	& $C^{x-y}_{xx} $&$= -0.061$ & $C^{x-y}_{xy} $&$= -1.999$ \\
	& $C^z_{xx} $&$= -0.554$~~~~ & $C^z_{xy} $&$= -0.253$ \\ \midrule
	$\bm{[}\si{\textbf{\electronvolt \angstrom}^{\textbf{2}}} \bm{]}$~~~~& $C^{x^2+y^2}_{00} $&$= -71.21$~ & $C^{x^2+y^2}_{z0} $&$= 56.30$ \\
	& $C^{xy}_{00} $&$= -137.1$ & $C^{xy}_{z0} $&$= 123.1$ \\
	& $C^{xz-yz}_{00} $&$= 1.52$ & $C^{xz - yz}_{z0} $&$= -1.49$ \\
	& $C^{z^2}_{00} $&$= -0.84$ & $C^{z^2}_{z0} $&$= -1.88$\\ 
	\bottomrule
\end{tabular}
\caption[]{Parameters of the $4 \times 4$ Hamiltonian of TaAs\sub{2} around $M$ up to second order.}
\label{tab:fit_parameters}
\end{table}

\section{Band structure topology}
In this section, we describe the band structure topology and the influence of magnetic field. First, we describe the topology in the absence of magnetic field for all four compounds. Then, we show that Weyl points appear under sufficient magnetic field. This result is shown first for the $\vec{k}\cdot\vec{p}$ model of TaAs\sub{2} derived in \cref{ssec:kdotp}, and then for a tight-binding model of NbSb\sub{2} derived from first-principles.
\subsection{Band structure topology without magnetic fields}

In the absence of magnetic field, there is no direct band gap closure in AB\sub{2} compounds. Since the valence bands thus form a well-defined manifold, they can be classified, just like insulators, according to the topology of these valence bands. Because time-reversal symmetry is fulfilled, a $\mathbb{Z}_2$ classification is possible.

\begin{figure}
	\includegraphics[width=\columnwidth]{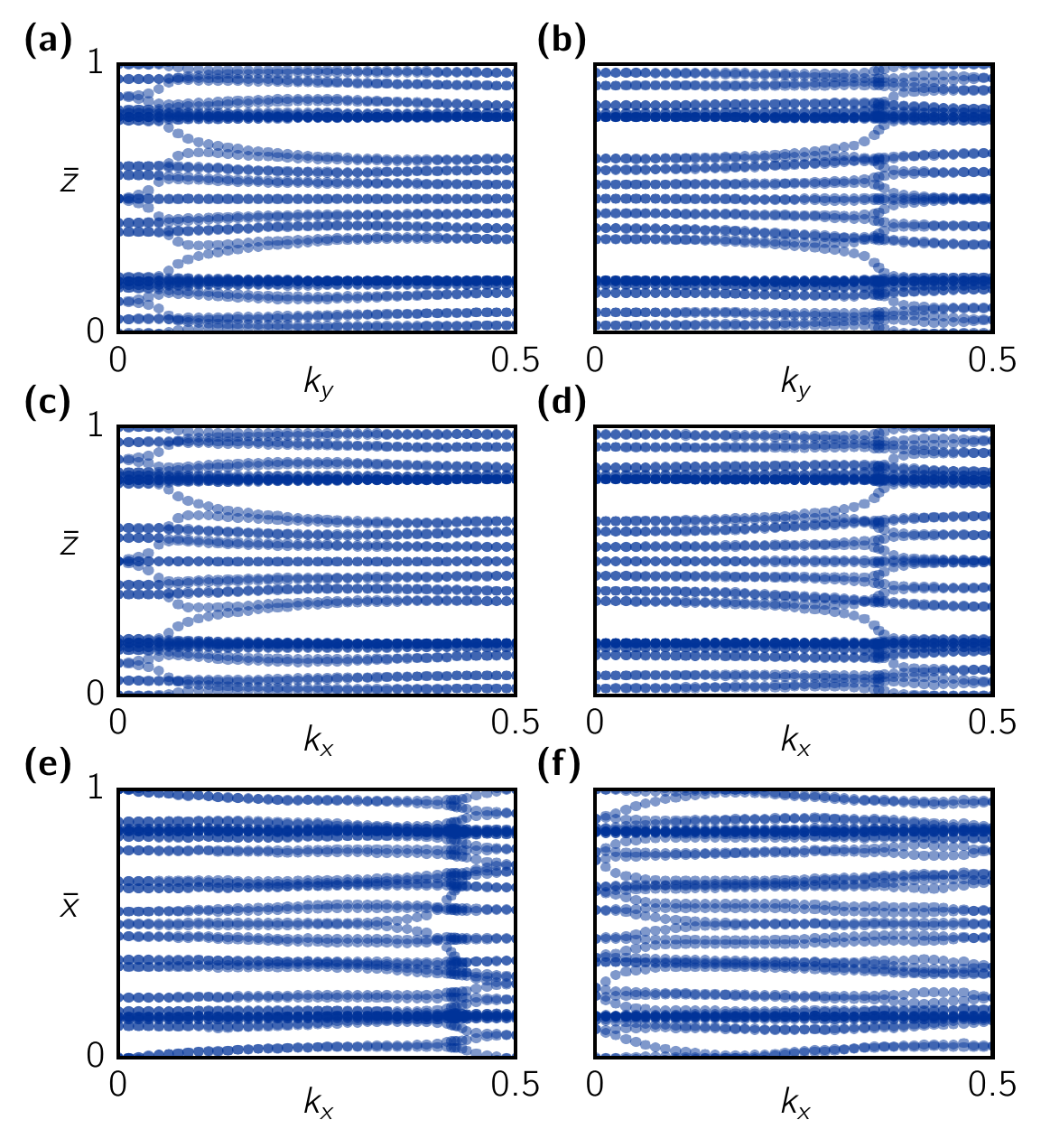}
	\caption[]{Wannier charge center evolution for the time-reversal invariant planes of TaAs\sub{2}. (a) $k_x=0$ (b) $k_x=0.5$ (c) $k_y=0$ (d) $k_y=0.5$ (e) $k_z=0$ (f) $k_z = 0.5$}
	\label{fig:z2_TaAs2}
\end{figure}

All compounds were found to be weak topological insulators, with $\mathbb{Z}_2$ indices $0;(111)$. That is, all time-reversal invariant planes $k_i=0,~0.5$ have a non-trivial $\mathbb{Z}_2$ index $\Delta=1$. This result was derived from first-principles using the \texttt{Z2Pack} code~\cite{Z2Pack}, and agrees with previous studies ~\cite{Luo-ScR16, Li-ARX16, Xu-PRB16}. The corresponding evolution of Wannier charge centers is shown, for the case of TaAs\sub{2}, in \cref{fig:z2_TaAs2}.

\begin{figure}
	\includegraphics[width=\columnwidth]{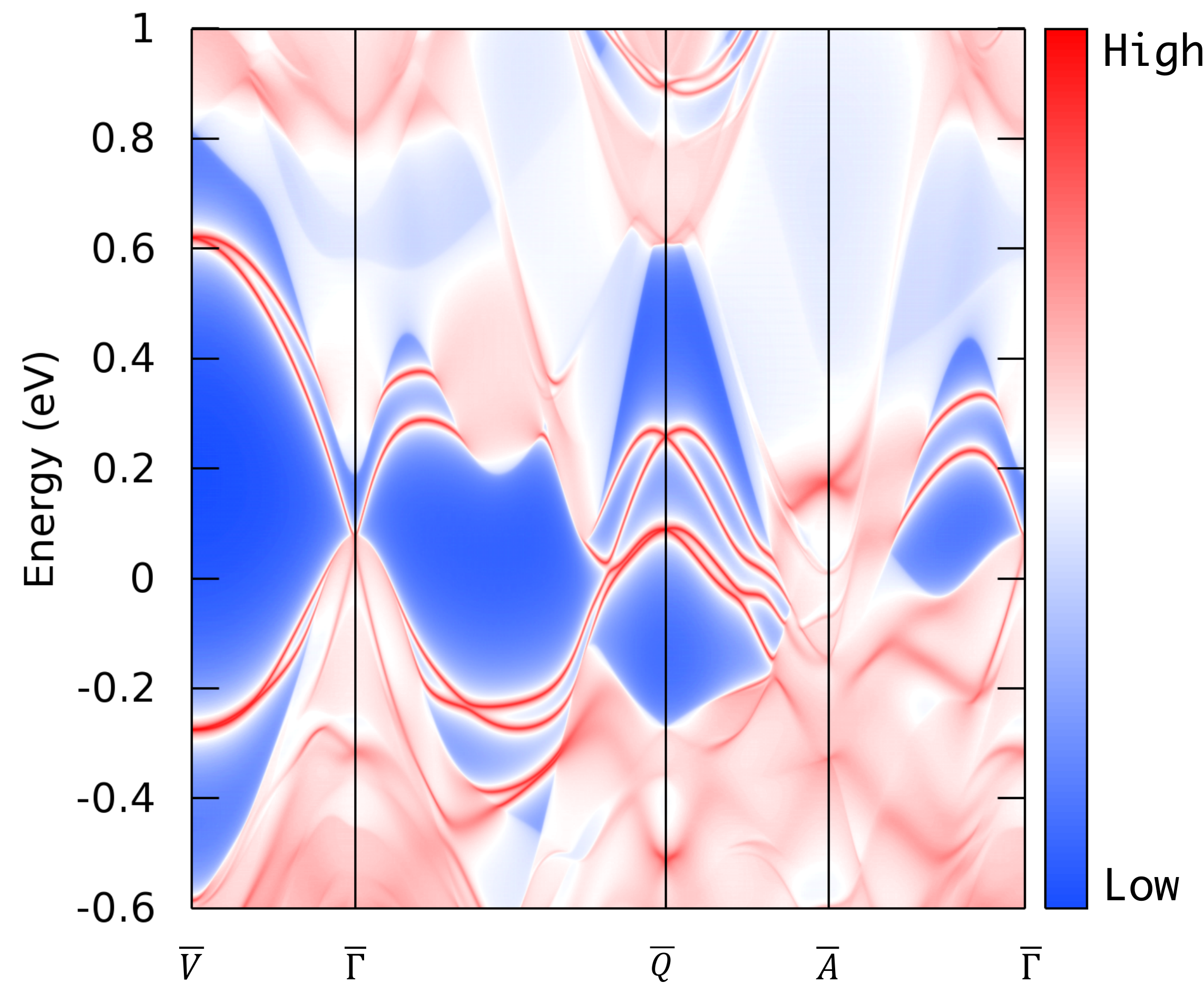}
	\caption[]{Surface density of states of TaAs\sub{2} on the 010 surface, along the k-point path shown if \cref{fig:uc_bz_TaAs2}.}
	\label{fig:TaAs2_surface_states}
\end{figure}

\Cref{fig:TaAs2_surface_states} shows the surface density of states for a slab of TaAs\sub{2}, with surfaces parallel to the mirror plane perpendicular to the cartesian $y$-axis (the light blue plane shown in \cref{fig:uc_bz_TaAs2}). The presence of topological surface states confirms the conclusion that the material is a weak topological insulator. The surface spectrum was calculated by the iterative Green's function~\cite{Sancho1985} which was implemented in WannierTools~\cite{wanniertools}. 

\subsection{Effect of Zeeman splitting on the $\mathbf{k \cdot p}$ model for TaAs\sub{2}} \label{ssec:zeeman_kdotp}

Here we study the effects of magnetic field on TaAs\sub{2} by adding a Zeeman splitting term to the \kdotp model derived in \cref{ssec:kdotp} (\cref{eqn:kdotp_hamilton}). The splitting term is given by
\begin{equation}\label{eqn:zeeman_kdotp}
\Delta \mathcal{H} = c_x \sigma_0 \otimes \sigma_y + c_y \sigma_0 \otimes \sigma_z + c_z \sigma_0 \otimes \sigma_x,
\end{equation}
where $c_i$ is the strength of the Zeeman splitting induced by the magnetic field in that direction, that is 
\begin{equation}
c_i = \sum_j g_{ij} \mu_\text{B}H_j.
\end{equation}

This assumes that the g-factor is equal for all bands. The limitations of this approximation are discussed in \cref{ssec:limitations}.

\subsubsection{Magnetic field along the rotation axis $\hat{y}$}
When magnetic field is applied along the rotation axis $\hat{y}$, the Zeeman term (\cref{eqn:zeeman_kdotp}) takes the form 
\begin{equation}
\Delta \mathcal{H} = c_y \sigma_0 \otimes \sigma_z .
\end{equation}
This term preserves all spatial symmetries of the system, breaking only time-reversal.

Along the $M - A$ line, the $C_{xx}$ and $C_{yy}$ contributions to the Hamiltonian vanish since $k_x^* = k_y^*$ and $k_z^*=0$. Consequently, the energy eigenvalues are given by
\begin{equation}\label{eqn:kdotp_zeeman_energy}
E(\vec{k}) = C_{00}(\vec{k}) \pm c_y \mp \sqrt{C_{xz}(\vec{k})^2 + C_{y0}(\vec{k})^2 + C_{z0}(\vec{k})^2} 
\end{equation}
The Zeeman term counteracts the original splitting (square root term), such that for sufficient magnetic field there will be a direct band gap closure. Away from the $M - A$ line, the band gap remains open, giving rise to a Weyl point.

\begin{figure}
	\includegraphics[width=\columnwidth]{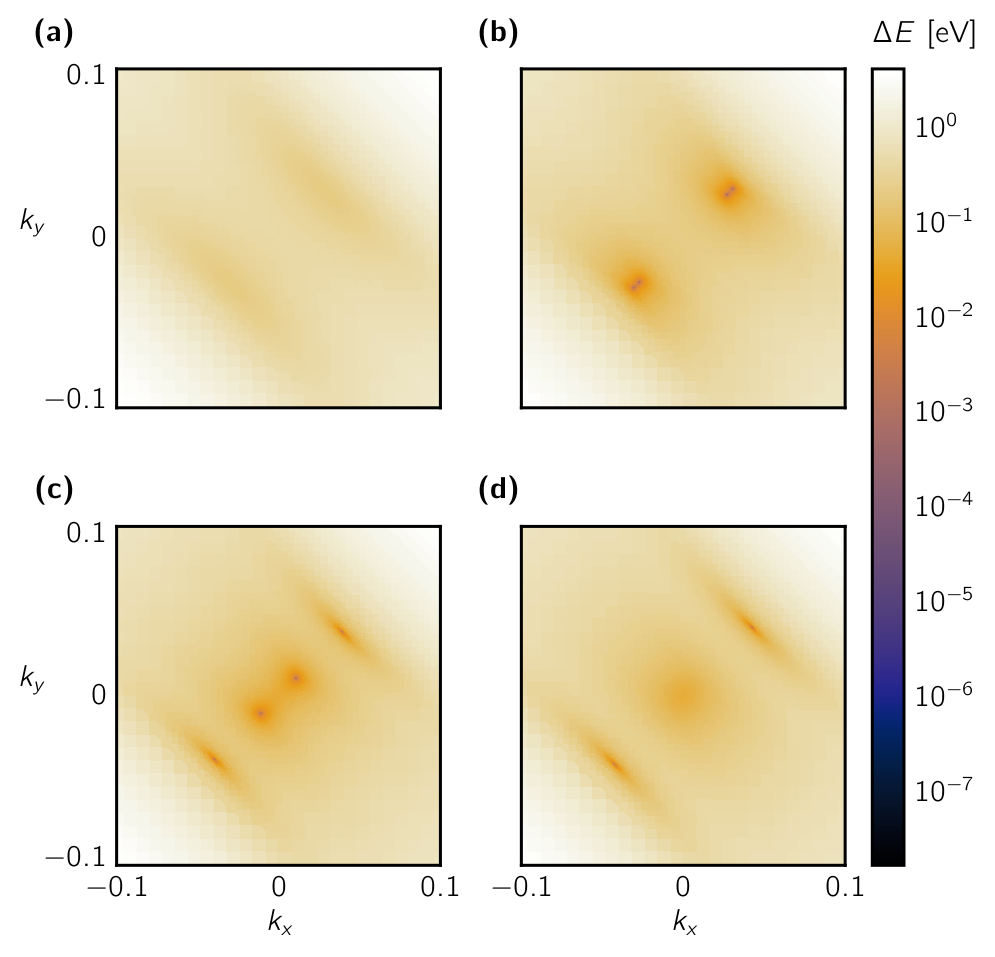}
	\caption[]{Band gap of TaAs\sub{2} in the $k_z^*=0$ plane for different values of the magnetic field in $y$-direction, calculated from the $\vec{k}\cdot \vec{p}$ model. A dark spot indicates the presence of a Weyl point.
		(a) No magnetic field. There are no Weyl points present 
		(b) $c_y = 0.11 \unit{eV}$. Two pairs of Weyl points have appeared on the $k_x = k_y$ line.
		(c) $c_y = 0.2 \unit{eV}$. The pair of Weyl points move further apart.
		(d) $c_y = 0.25 \unit{eV}$. One pair of Weyl points has annihilated at $M$, leaving two Weyl points.
	}
	\label{fig:phase_transition}
\end{figure}

When the Zeeman splitting is gradually switched on, two pairs of Weyl points form at about $c_y = 0.11 \unit{eV}$. Increasing the Zeeman splitting leads to a separation between the two nodes in a pair, with one node each moving towards the $M$-point. Finally, at $c_y \approx 0.25 \unit{eV}$, these two nodes meet at $M$ and annihilate. This process is shown in \cref{fig:phase_transition}.

\begin{figure}
\includegraphics[width=\columnwidth]{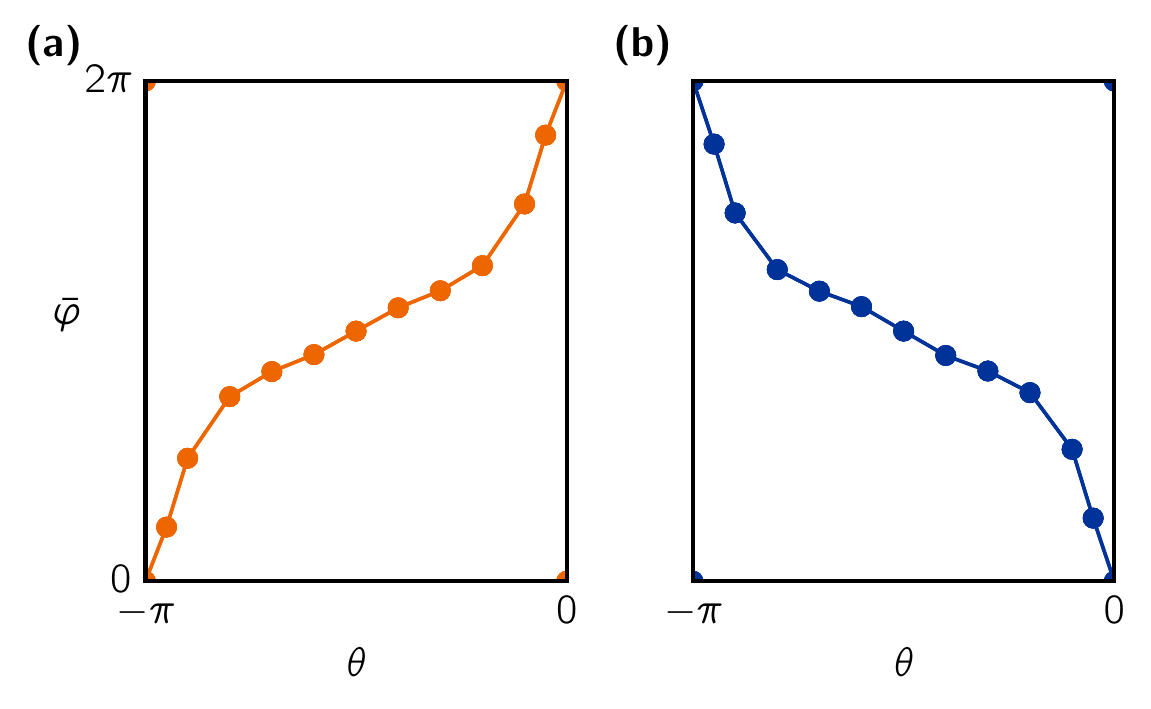}
\caption[]{Evolution of the sum of HWCC on spheres surrounding the Weyl points at $c_y=0.12 \unit{eV}$ splitting. (a) Weyl point at $\vec{k}=(0.5247, 0.5247, 0.5)$, having positive chirality $C=+1$ (b) Weyl point at $\vec{k}=(0.53258, 0.53258, 0.5)$ with negative chirality $C=-1$}
\label{fig:TaAs2_chern}
\end{figure}

The existence of these Weyl points was confirmed by verifying that the nodes are a source or sink of Berry curvature. For this purpose, the Chern number of spheres surrounding the points was calculated by tracking hybrid Wannier charge centers (HWCC) on loops around the sphere~\cite{Soluyanov-Nat15, Wang-PRL16, Autes-PRL16, Z2Pack}, using the \texttt{Z2Pack} software ~\cite{Z2Pack}. \Cref{fig:TaAs2_chern} shows the evolution of the sum of HWCC for two of four nodes found at $c_y=0.12 \unit{eV}$, demonstrating that the two points are Weyl nodes of opposite chirality.

\subsubsection{General magnetic field direction}
Finally, the effects of a magnetic field in a general direction were studied. It turns out that, even though such a field breaks the spatial symmetries of the system, Weyl nodes still appear under a strong enough magnetic field. When magnetic field is applied in $\hat{x}$ or $\hat{z}$ - direction, a single pair of Weyl points emerges from the $M$ point. These Weyl nodes are located on the $k_x=-k_y$ plane, as shown in \cref{tab:weyl_pos_TaAs2}.

\begin{table}[h]
\begin{tabular}{@{}llr@{}}
\toprule 
\textbf{Splitting} $\bm{[}\si{\textbf{\electronvolt}} \bm{]}$~~ & \textbf{Weyl position} $\vec{k^*}$ & \textbf{Chirality}\\ \midrule
$c_x = 0.225$ & $(-0.0042, 0.0042, 0.00093)$~~ & $-1$\\
& $(0.0042, -0.0042, -0.00093)$~~ & $1$ \\\midrule
$c_x = 0.25$ & $(-0.025, 0.025, 0.0054)$ & $-1$\\
& $(0.025, -0.025, -0.0054)$ & $1$\\ \midrule
$c_x = 0.3$ & $(-0.044, 0.044, 0.0098)$ & $-1$ \\
& $(0.044, -0.044, -0.0098)$ & $1$ \\ \midrule[.08em]
$c_z = 0.225$ & $(0.0011, -0.0011, -0.018)$~~ & $-1$ \\
& $(-0.0011, 0.0011, 0.018)$~~ & $1$ \\ \midrule
$c_z = 0.25$ & $(0.0066, -0.0066, -0.11)$ & $-1$ \\
& $(-0.0066, 0.0066, 0.11)$ & $1$ \\ \midrule
$c_z = 0.3$ & $(0.012, -0.012, -0.18)$ & $-1$ \\
& $(-0.012, 0.012, 0.18)$ & $1$\\ \bottomrule
\end{tabular}
\caption[]{Position $\vec{k^*} = \vec{k} - M$ (in reduced coordinates) and chirality of Weyl points for Zeeman splittings in $\hat{x}$ and $\hat{z}$-direction.}
\label{tab:weyl_pos_TaAs2}
\end{table}

\begin{figure}
\includegraphics[width=\columnwidth]{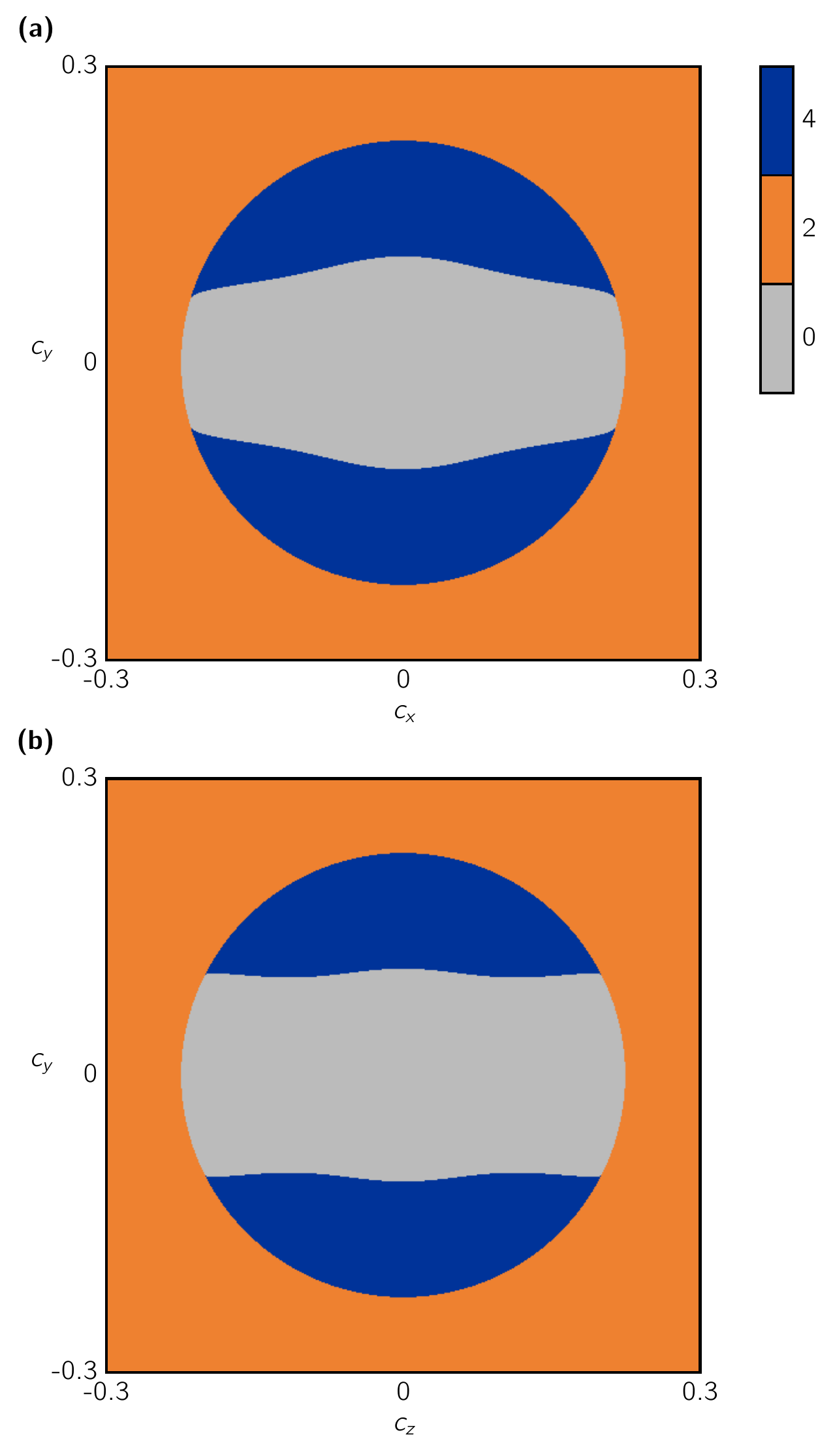}
\caption[]{Phase diagram showing the number of Weyl points in the $\vec{k}\cdot \vec{p}$ model of TaAs\sub{2} as a function of the Zeeman splitting (in $\unit{eV}$).}
\label{fig:phases_TaAs2}
\end{figure}

\Cref{fig:phases_TaAs2} shows the number of Weyl points as a function of the Zeeman splitting.
To obtain this phase diagram, candidate Weyl points were identified using a quasi Newton algorithm to find minima in the band gap (using \texttt{scipy.optimize.minimize}~\cite{scipy}), for different initial guesses. In a second step, the Chern number on a small sphere (radius $10^{-4} \unit{\AA^{-1}}$) surrounding the candidate points was evaluated (using \texttt{Z2Pack}~\cite{Z2Pack}), keeping only points with a non-zero Chern number. Finally, duplicate points were eliminated by checking whether two points lie within the diameter of the sphere of one another.

\subsection{Effect of Zeeman splitting on the tight-binding model for NbSb\sub{2}}
Having studied the effects of Zeeman splitting on the \kdotp model for TaAs\sub{2}, we now study a more realistic tight-binding model for NbSb\sub{2}, derived from a first-principles calculation with hybrid functionals using the Wannier90 code~\cite{wannier90, wannier90_14}. NbSb\sub{2} was chosen because it has the smallest direct band gap of the four materials, making it the most promising candidate for hosting Weyl points at realistic magnetic field strength.

The Zeeman splitting for this model can again be expressed by adding the corresponding terms to the Hamiltonian
\begin{equation}
\Delta \mathcal{H} = c_x \sigma_x \otimes \mathbb{I}_{22 \times 22} + c_y \sigma_y \otimes \mathbb{I}_{22 \times 22} + c_z \sigma_z \otimes \mathbb{I}_{22 \times 22},
\end{equation}
where the change in the splitting terms (compared to \cref{eqn:zeeman_kdotp}) is due to the different orbital basis used for the tight-binding model. We search for Weyl points between the last valence band and the first conduction band.

\begin{figure}
	\includegraphics[width=\columnwidth]{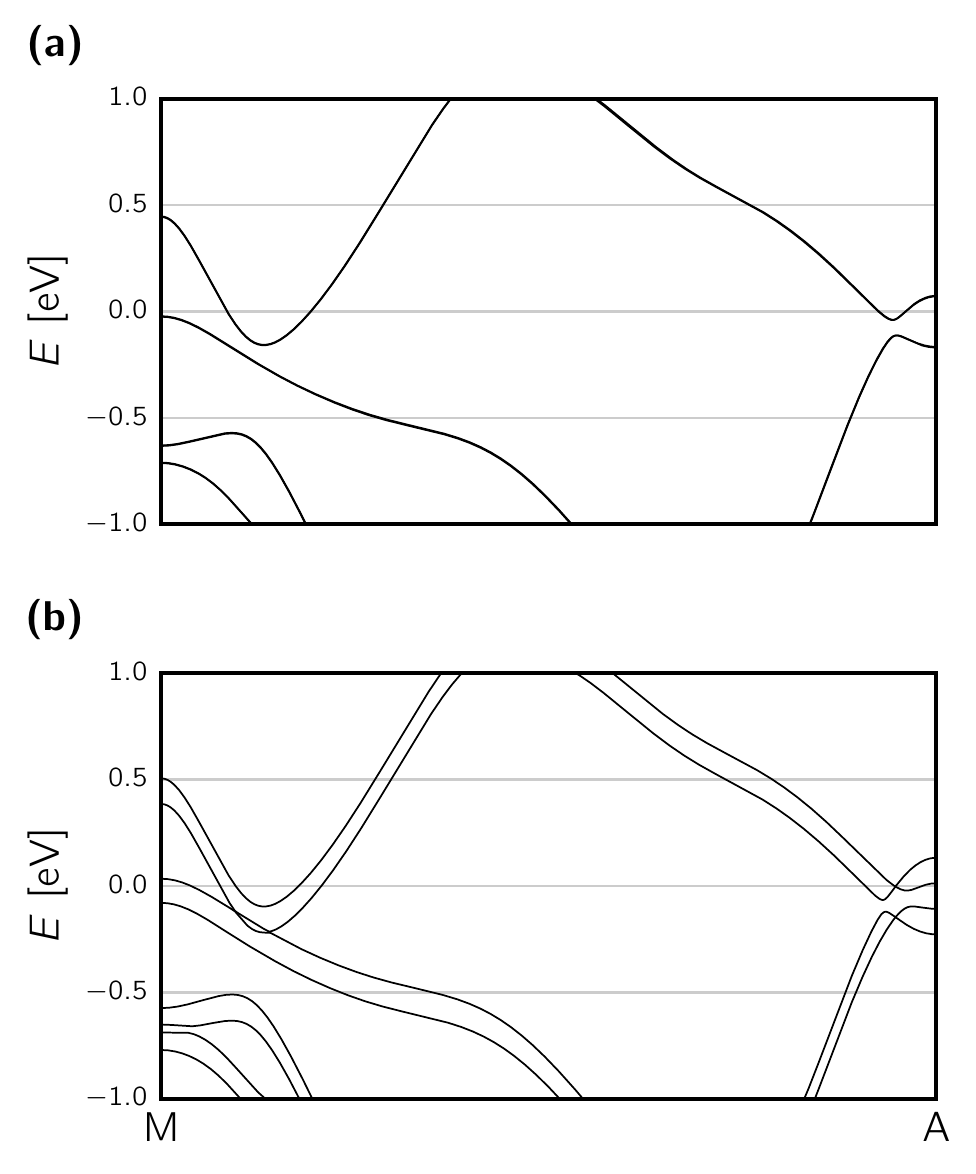}
	\caption[]{Band structure of the tight-binding model for NbSb\sub{2} along the $M - A$ line. (a) Without Zeeman splitting. (b) With $c_y = 0.06 \unit{eV}$ Zeeman splitting.}      
	\label{fig:NbSb2_splitting_bands}
\end{figure}

First we study the effect of applying a magnetic field in the $y$-direction. \Cref{fig:NbSb2_splitting_bands} shows the effect of this splitting along the $M - A$ line. For $c_y \approx 0.06 \unit{eV}$, two pairs of Weyl points appear close to the $M - A$ line. The reason these points are not exactly on the line is because the crystal symmetry is broken when constructing the Wannier-based tight-binding model~\cite{wannier90,wannier90_14}. Apart from this numerical difference, this effect is analogous to the case of the \kdotp model for TaAs\sub{2}, where the two pairs of Weyl points appeared at $c_y = 0.11 \unit{eV}$. 

\Cref{tab:NbSb2_weyl} shows the Weyl point positions, chirality and type for selected values of the Zeeman splitting. It shows that Weyl points appear even at smaller values of $c_y$ away from the $M-A$ line. This is a crucial difference to the \kdotp model which is valid only near the $M$ point. Furthermore, all Weyl points found for these splitting values are of type II~\cite{Soluyanov-Nat15}. Type - II Weyl points have a tilted energy spectrum, making their Fermi surface open instead of point-like. As a consequence, their chiral anomaly -- and their effect on magnetoresistance -- is expected to be anisotropic.

\begin{table}
\begin{tabular}{@{}llrr@{}}
	\toprule 
	\textbf{Split.} $\bm{[}\si{\textbf{\electronvolt}} \bm{]}$~~ & \textbf{Position} $\vec{k}$ & \textbf{Chir.} & \textbf{Type}\\ \midrule
	$c_x = 0.045$ & $(0.4393, 0.4460, 0.5004)$ & $+1$ & II \\
	& $(0.4359, 0.4444, 0.5026)$ & $-1$ & II \\
	& $(0.5641, 0.5556, 0.4974)$ & $+1$ & II \\
	& $(0.5607, 0.5540, 0.4996)$ & $-1$ & II \\\midrule[.08em]
	$c_y = 0.03$ & $(0.3670, 0.5141, 0.0977)$~~ & $+1$ & II\\
	& $(0.3655, 0.5142, 0.1004)$ & $-1$ & II\\ 
	& $(0.6345, 0.4858, 0.8997)$ & $+1$ & II\\ 
	& $(0.6330, 0.4858, 0.9023)$ & $-1$ & II\\ \midrule
	$c_y = 0.04$ & $(0.3724, 0.5116, 0.0890)$ & $+1$ & II \\
	& $(0.3627, 0.5135, 0.1055)$ & $-1$ & II \\
	& $(0.6373, 0.4865, 0.8945)$ & $+1$ & II \\
	& $(0.6276, 0.4884, 0.9110)$ & $-1$ & II \\
	& $(0.9028, 0.0340, 0.5451)$ & $+1$ & II \\
	& $(0.9018, 0.0354, 0.5390)$ & $-1$ & II \\
	& $(0.0982, 0.9646, 0.4610)$ & $+1$ & II \\
	& $(0.0974, 0.9658, 0.4545)$ & $-1$ & II \\ \midrule
	$c_y = 0.06$
	& $(0.3791, 0.5068, 0.0775)$ & $+1$ & II \\
	& $(0.3592, 0.5131, 0.1108)$ & $-1$ & II \\
	& $(0.6407, 0.4869, 0.8892)$ & $+1$ & II \\
	& $(0.6211, 0.4929, 0.9222)$ & $-1$ & II \\
	& $(0.9033, 0.0328, 0.5532)$ & $+1$ & II \\
	& $(0.9006, 0.0364, 0.5314)$ & $-1$ & II \\
	& $(0.0994, 0.9636, 0.4686)$ & $+1$ & II \\
	& $(0.0968, 0.9671, 0.4467)$ & $-1$ & II \\
	& $(0.4493, 0.4555, 0.5031)$ & $+1$ & II \\
	& $(0.4309, 0.4320, 0.4825)$ & $-1$ & II \\
	& $(0.5691, 0.5680, 0.5175)$ & $+1$ & II \\
	& $(0.5507, 0.5445, 0.4969)$ & $-1$ & II \\
	\midrule[.08em]
	$c_z = 0.0475$
	& $(0.4494, 0.4384, 0.4853)$ & $+1$ & II \\
	& $(0.4420, 0.4366, 0.4816)$ & $-1$ & II \\
	& $(0.5580, 0.5634, 0.5184)$ & $+1$ & II \\
	& $(0.5506, 0.5616, 0.5147)$ & $-1$ & II \\
	\bottomrule
\end{tabular}
\caption[]{Weyl point positions (in reduced coordinates), chirality and type for different values of the Zeeman splitting in the tight-binding model for NbSb\sub{2}.}
\label{tab:NbSb2_weyl}
\end{table}

\begin{figure*}

\begin{minipage}[t]{\columnwidth}
\includegraphics[width=0.970588235\columnwidth]{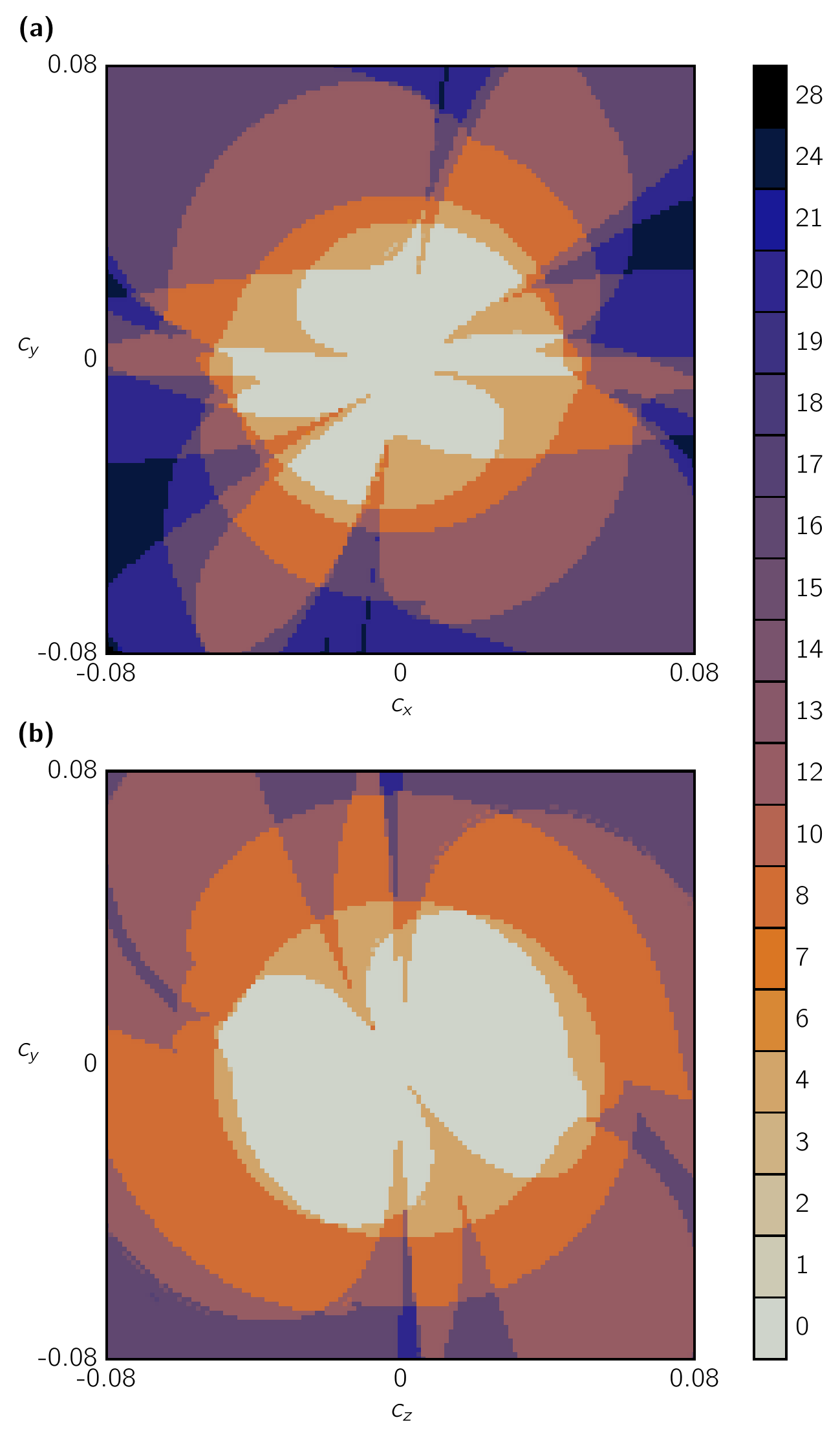}
\caption[]{Phase diagram showing the number of Weyl points as a function of Zeeman splitting (in \si{\electronvolt}) for the tight-binding model for NbSb\sub{2}. }
\label{fig:NbSb2_phase_diagram}
\end{minipage}\hfill
\begin{minipage}[t]{\columnwidth}
\includegraphics[width=\columnwidth]{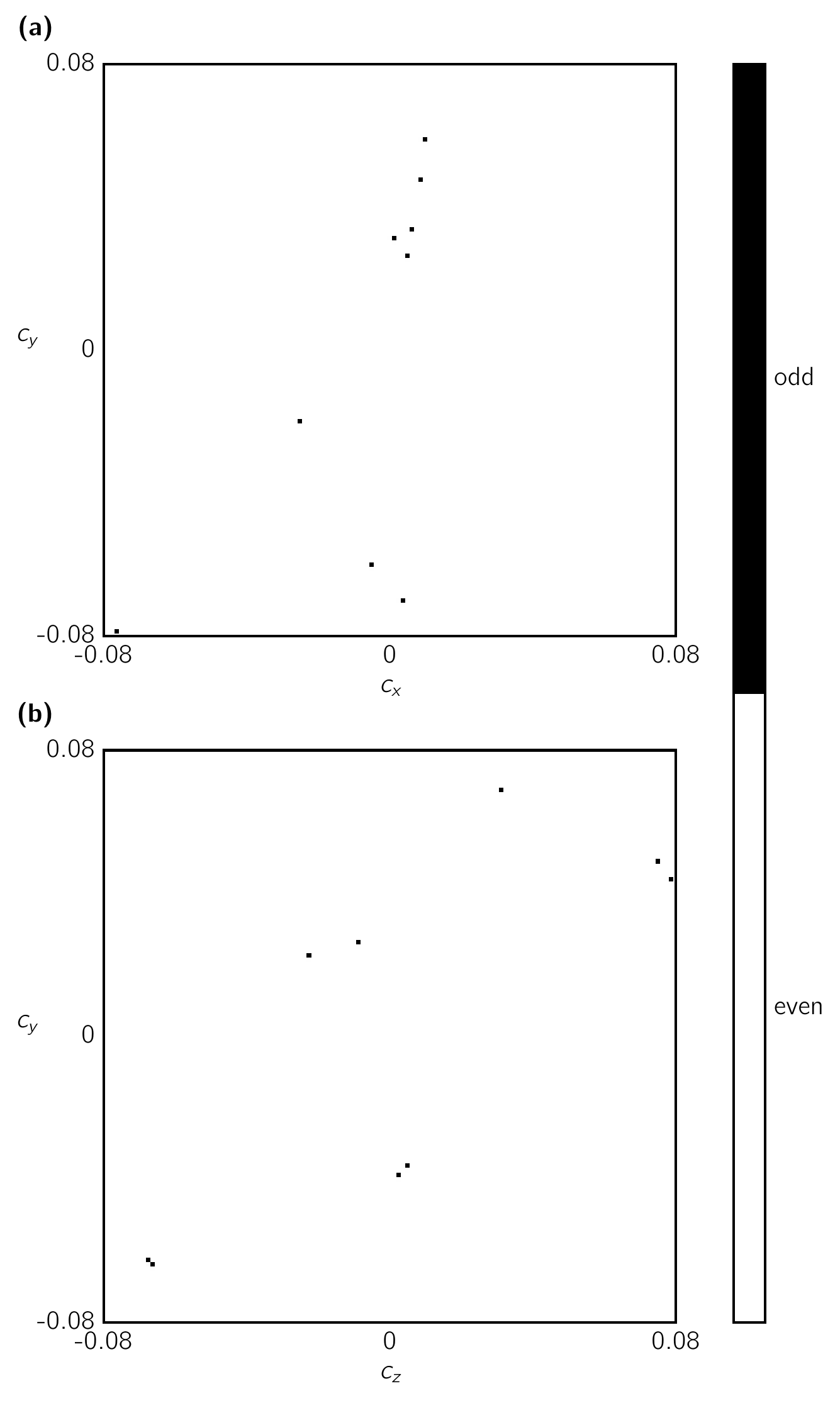}
\caption[]{Phase diagram showing whether the number of Weyl points shown in \cref{fig:NbSb2_phase_diagram} is even (physically possible) and odd (non-physical). The odd phases are a relict of the numerical evaluation of the phase.}
\label{fig:NbSb2_odd_phases}
\end{minipage}
\end{figure*}

Finally, a phase diagram showing the number of Weyl points as a function of magnetic field was calculated (see \cref{fig:NbSb2_phase_diagram}). Unlike for the \kdotp model, the number of Weyl points keeps increasing when the applied Zeeman term grows stronger. Again, the reason for this difference is that Weyl points also form far away from the $M$ point, where the \kdotp approximation is no longer applicable. 

For some values of the splitting, the phase diagram shows an odd number of Weyl points, which is physically impossible. The reason for this is that the numerical procedure used to identify the number of Weyl points may not find a Weyl point if it is too close to another Weyl point. Since this problem occurs only rarely (see \cref{fig:NbSb2_odd_phases}), the phase diagram is still valid overall. Also, the procedure ensures that no Weyl point can be counted twice, so the phase diagram represents a lower limit for the real number of Weyl points. Thus, the general result that the number of Weyl points increases with stronger Zeeman splitting remains valid.

\subsection{Limitations of the model for magnetic field} \label{ssec:limitations}
In the previous sections the effect of magnetic field was modeled by applying a Zeeman splitting to the model Hamiltonian. The discussion was simplified by assuming that the g-factor is equal for all energy bands, and independent of $\vec{k}$. Here we discuss how the results might change if this assumption is not made.

If the g-factor is $k$-dependent, but still the same for all energy bands, the results above will change quantitatively, but not qualitatively. The reason for this is that a Weyl node that appears at a specific $k$-point will still be there, but for a different magnetic field. That is, the order in which the Weyl nodes at different $k$-points appear might change, but not the overall picture that there is an increasing number of Weyl points with stronger magnetic field. 

The same is true if the g-factor varies for different energy bands, as long as the sign of the g-factor remains the same. Because the appearance of Weyl points is due to the relative Zeeman splitting between the last valence and first electron bands, it does not matter how much the splitting on each band contributes. 

If the g-factors in the relevant bands have opposite sign however, there is a qualitative change in the behavior. This is illustrated in the following with the example of the \kdotp model of TaAs\sub{2} discussed in \cref{ssec:kdotp,ssec:zeeman_kdotp}. To account for the opposite sign of the g-factor for valence and conduction bands, the Zeeman splitting term (\cref{eqn:zeeman_kdotp}) is changed to
\begin{equation}\label{eqn:zeeman_kdotp_inverted}
\Delta \mathcal{H} = c_x \sigma_z \otimes \sigma_y + c_y \sigma_z \otimes \sigma_z + c_z \sigma_z \otimes \sigma_x.
\end{equation}

\begin{figure}
\vspace*{0.1cm}
\includegraphics[width=\columnwidth]{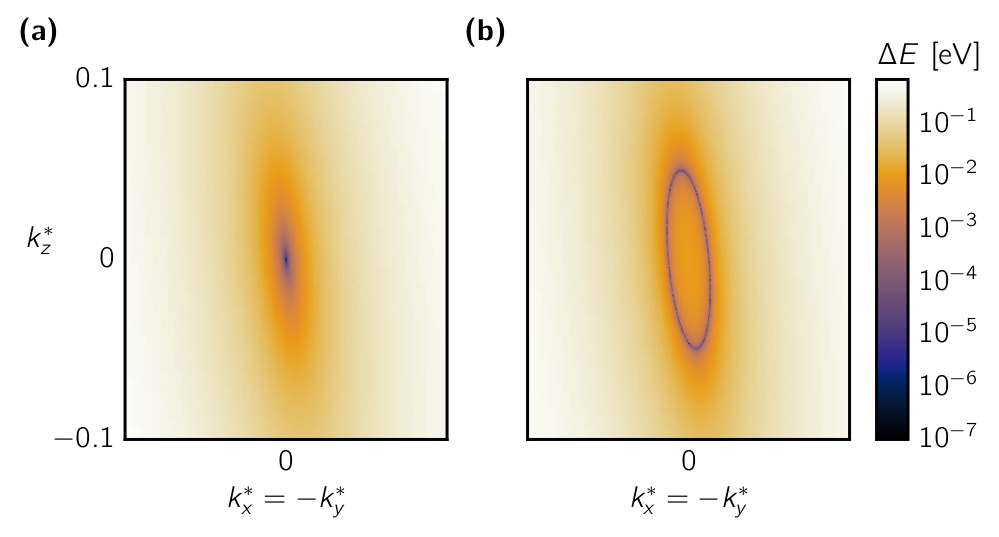}
\caption[]{Band gap of TaAs\sub{2} on the mirror plane with Zeeman splitting as given in \cref{eqn:zeeman_kdotp_inverted}. (a) At $c_y\approx 0.2242 \unit{eV}$, a nodal line appears at the $M$-point. (b) The nodal line expands for stronger splitting ($c_y = 0.23 \unit{eV}$). }
\label{fig:TaAs2_nodal_line}
\end{figure}

With $c_y$ splitting, the energy bands on the mirror plane is then given by
\begin{equation}
E(\vec{k}) = C_{00}(\vec{k}) \pm c_y \mp \sqrt{C_{xx}(\vec{k})^2 + C_{xy}(\vec{k})^2 + C_{z0}(\vec{k})^2 }.
\end{equation}
As in \cref{eqn:kdotp_zeeman_energy}, the Zeeman term counteracts the original splitting. The difference to the previous case is that this equation holds on an entire plane in reciprocal space instead of just a line. As consequence, we can expect the appearance of a nodal line with sufficient Zeeman splitting. Indeed, a nodal line appears for $c_y \gtrsim 0.2242 \unit{eV}$, as shown in \cref{fig:TaAs2_nodal_line}. The Berry phase on a closed path around this nodal line was calculated to be $\pi$, using the \texttt{Z2Pack}~\cite{Z2Pack} software. This verifies the topological nature of the nodal line.

In conclusion, the qualitative result obtained above remains intact when the g-factors are assumed to be $k$-dependent and different for valence and conduction bands, as long as they keep the same sign. A more adequate model of the magnetic field is needed to establish the exact qualitative and quantitative nature of the topological phases with applied magnetic field. The current results indicate that Weyl nodes will appear at least for some directions of magnetic field. 

\section{Conclusions}

We studied the topological phase of transition metal dipnictides of the type AB\sub{2} (A $\in$ \{Ta, Nb\}, B $\in$ \{As, Sb\}), with and without external magnetic field. 
In the absence of magnetic field, we found -- in accordance with previous results~\cite{Luo-ScR16, Li-ARX16} -- that these materials can be classified as weak topological insulators despite having an indirect band gap closure. 

The effect of a magnetic field was studied by applying a Zeeman splitting. We found that Weyl points can appear. We showed this result first from theoretical considerations on a four-band \kdotp model, and numerically on a \kdotp model of TaAs\sub{2} and a tight-binding model of NbSb\sub{2}. In the tight-binding model, we found the number of Weyl points to be increasing with growing magnetic field. For specific values of the Zeeman splitting, the type of the Weyl points in the tight-binding model was studied, and they were all found to be of type II.

The appearance of such field-induced Weyl points could help explain the reduced or negative magneto-resistivity in these materials. However, it is unclear whether the Weyl points studied here appear at a magnetic field that is realistic to observe in experiments. 
Further studies, in particular to obtain a realistic $g$-factor and more reliable data for the direct band gap, are required to accurately estimate the required magnetic field. 
Furthermore, it is known that modeling a strong magnetic field with only Zeeman splitting is not sufficient, and a more accurate model should be considered. Finally, the effect of these Weyl points on the magnetoresistance should be calculated. This is influenced by the orientation of the type-II Weyl points, and their distance from the Fermi level.

Consequently, there are three open questions which require further investigation: First, whether the appearance of field-induced Weyl points is realistic in these AB\sub{2} compounds. Second, if these Weyl points do appear, whether they alone are responsible for the experimentally observed behavior of magneto-resistance or if there are other effects. Finally, whether there are other compounds which contain the same kind of field-induced Weyl points, possibly appearing already at weaker magnetic field.

\section{Acknowledgments} 
We would like to thank D. Rodic and M. K\"onz for helpful discussions. The authors were supported by Microsoft Research, and the Swiss National Science Foundation through  the  National  Competence  Centers  in  Research MARVEL and QSIT.


\bibliography{paper}


\end{document}